\documentclass[a4paper, 10pt, conference]{ieeeconf}      % Use this line for a4 paper

\IEEEoverridecommandlockouts                              % This command is only needed if 
\usepackage{graphics} % for pdf, bitmapped graphics files
\usepackage{epsfig} % for postscript graphics files
\usepackage{mathptmx} % assumes new font selection scheme installed
\usepackage{times} % assumes new font selection scheme installed
\usepackage{amsmath} % assumes amsmath package installed
\usepackage{amssymb}  % assumes amsmath package installed

\usepackage{graphicx}           	% Include this line if your 
\usepackage{epstopdf}           	%to include eps figure in the latex file
\usepackage{subfloat}				%to include subfigures
\usepackage{subfigure}
\usepackage{url}
\usepackage{tikz}
\usepackage{verbatim}

\def\angle{0}
\def\radius{3}
\def\cyclelist{{"green","red"}}
\newcount\cyclecount \cyclecount=-1
\newcount\ind \ind=-1

\newcommand{\V}{\mathcal{V}}
\newcommand{\E}{\mathcal{E}}
\newcommand{\G}{\mathcal{G}}
\newcommand{\N}{\mathcal{N}}
\newtheorem{definition}{Definition}

\newtheorem{theorem}{Theorem}

\newtheorem{problem}{Problem}
\newcommand{\figurename}[1]{Figure~{#1}}

\title{\LARGE \bf
Social network analysis of electric vehicles adoption: \\a data-based approach
\thanks{Preprint submitted to the 1st IEEE International Conference on Human-Machine Systems (ICHMS) 2020.}}

\author{V. Breschi$^{a}$, M. Tanelli$^{a,}$$^{b}$, C. Ravazzi$^{b}$, S. Strada$^{a}$ and F. Dabbene$^{b}$% <-this % stops a space
	\thanks{$^{a)}$ Dipartimento di Elettronica, Informazione e Bioingegneria, Politecnico di Milano, Piazza Leonardo da Vinci 32, 20133 Milano, Italy.}%    
		\thanks{$^{b)}$ Istituto di Elettronica e Ingegneria dell'Informazione e delle Telecomunicazioni - IEIIT CNR
Corso Duca degli Abruzzi 24, 10129 Torino}%  
}  

\begin{document}

\maketitle
\thispagestyle{empty}
\pagestyle{empty}

%%%%%%%%%%%%%%%%%%%%%%%%%%%%%%%%%%%%%%%%%%%%%%%%%%%%%%%%%%%%%%%%%%%%%%%%%%%%%%%%
\begin{abstract}
Mobility is undergoing dramatic transformations. Especially in the context of urban areas, several significant changes are underway, driven by both new mobility needs and environmental concerns. The most mature one, which still is struggling to affirm itself is the process of the adoption of Electric Vehicles (EVs), thus switching from fuel-based to battery-powered propulsion technologies. Many social and economic barriers have proved to play a crucial role in this process, ranging from level of education, environmental awareness, age and census. This work aims at contributing to the study of this adoption process through a data-based lens, using real mobility patterns to setup a social-network analysis to model the spread of consensus among neighboring people that can enable the switch to EVs. In particular, we build the network topology using proximity measures that emerge from the analysis of real trips, and the initial disposition of the single agents towards the EV technology is inferred from their real mobility patterns. Based on this network, a cascade adoption model is simulated to investigate the dynamics of the adoption process, and an incentive scheme is designed to show how different policies can contribute to the opinion diffusion over time on the network.

\end{abstract}

%%%%%%%%%%%%%%%%%%%%%%%%%%%%%%%%%%%%%%%%%%%%%%%%%%%%%%%%%%%%%%%%%%%%%%%%%%%%%%%%
\section{Introduction}

The way in which people are moving, especially in urban areas, is experiencing several major paradigm shifts. These are necessary to answer the needs of a growing number of people living that are aggregating to live in the so-called \textit{mega-cities} (more that 60\% of the world population is expected to leave in a mega-city by 2030), while being sustainable and energy efficient, \cite{docherty2018governance}.

Two of the main the directions of change that mobility is facing are the following: (i) from internal-combustion-engine (ICE) to battery-operated electric vehicles (BEV): oil-based propulsion needs to be abandoned and substituted by fully electric or at most hybrid propulsion for long-range vehicles. In fact, transportation in its current form is responsible for the emission of over a quarter of all greenhouse gases (GHG) in Europe; in particular, passenger cars alone account for 41\% of such emissions (11\% of the total), \cite{grelier_2018}. A shift towards more eco-friendly means of mobility such as EVs could substantially contribute to lowering the amount of GHG emissions caused by road transport. This holds especially true when we consider that EVs are able to convert approximately 60\% of the electrical energy from the grid to power at the wheels, while conventional ICE vehicles only convert approximately 20\% of the fuel energy to power at the wheels, \cite{efficiency2017, needell2016potential}; (ii) a transition from ownership to usership: vehicle-sharing systems seem one of the most promising way for sustainable mobility in urban areas. With more than half of the world’s population that will soon live in metropolitan areas, the current private-vehicle-based mobility would lead to unmanageable traffic density, energy consumption, pollution, and congestion. Sustainable and shared mobility solutions in the context of highly-densely populated areas can also help to fulfill the mobility needs of medium to low-income people, as they lower the ownership costs of an EV. 

From the above discussion, it is clear that mobility as a whole will be more and more developed intertwining technological, social and economic needs. As such, the new, sustainable and shared mobility models cannot but be designed with human-centered principles, so as to embed in their functioning interactions with the users, and to manage their adaptation and evolution not only on the basis of technological needs, but also of social ones, \cite{barr2014smarter}. In the two examples above, a fundamental mindset shift must take place in the final user to allow a significant penetration of the two new mobility models, \textit{i.e.}, electric and shared vehicles.  In fact, as far as EVs are concerned, despite the sales growth of the recent years (registering a year-on-year market growth of 63\% in 2018, \cite{globalEV2019}) the penetration of electric cars is still limited to less than 1\% of the global car fleet, \cite{globalEV2019}. The greatest barriers to widespread EV adoption are the purchase price, the driving range and the charging availability, \cite{mckinsey_2017}. Even though new EVs have considerable driving ranges, still the so-called \textit{range anxiety}, i.e., the fear that users have to remain without battery charge while still on the road, is still a strong inhibiting factor in the EV diffusion. Also in the shift from ownership to usership strong social changes must be in place. Up to now owning a (possibly high-end) car is a strong social marker associated to high-status of individuals. Removing such social link to ownership is one of the main keys to allow the spread of the new model. 
A sensible framework to mix technological and social variables for analyzing the aforementioned adoption processes is that of opinion dynamics, which allows modelling the formation of beliefs, the aggregation of opinions and their dynamic evolution in a hyper-connected society, all within a quantitative and rigorous framework, see \textit{e.g.}, \cite{6160999, frasca2013gossips,friedkin2016network,lorenz2007continuous} and references therein.

In this work, we concentrate on the analysis of the EV-adoption process, and we use as a basis for modelling the mobility habits a large data-set of vehicle-based measurements, obtained via telematic e-Boxes, that followed thousands of users in a large Italian province for one year in their daily trips. These data are used to extract real mobility patterns that quantitatively describe the feasibility, for each single vehicle, to be seamlessly replaced by an EV. Based on this data-based scenario, we build a social network description of the considered users based on a proximity measure, and then analyze the EV-adoption process under different perspectives. Further, we show how incentives policies can sustain and speed this adoption process.
To the best of our knowledge, this is one of the very first contribution that uses the opinion dynamic framework in a data-driven way, which we believe is essential to consider the impact of social variables while starting from a realistic picture of the current needs of today’s mobility. We believe that only a meaningful mix of the two can yield sensible answers that allow taking into account users characteristics in the design of new mobility services.

The structure of the paper is as follows. Section \ref{Sec:cascading_model} introduces the graph-based model based on which the network topology is defined. Then, Section \ref{sec:adoption} illustrates the data-based analysis of the EV-adoption process, and how from that analysis we extracted the needed information for setting up the social network within an opinion dynamics framework. Based on it, Section \ref{sec:anal} analyses the diffusion of EV adoption under different perspectives, and Section \ref{sec:incentive} discusses the effects of possible incentive policies to support a faster adoption spread.

\section{Cascading network model}\label{Sec:cascading_model}
For modelling the EV-adoption process of interest, we need to build the influence network that describes the interaction between users based on proximity. To do that, we will now describe the cascading network model that will be the basis for the subsequent analysis.
To start with, let us consider a simple and undirected graph $\G=(\V,\E)$ where $\V$ is the set of nodes representing the agents and $\E\subseteq\V\times\V$ is the set of edges.
The assertion $(u,v)\in\E$ implies that agent $v$ is influenced by agent $u$ and, since the graph is undirected, also $(v,u)\in\E$. We denote the set of in-neighbors of a node $v\in\V$ by $\N_v=\{w\in\V:(w,v)\in\E\}$. A graph $\G$ is called strongly connected if there is a path from each vertex in the graph to every other vertex.

We consider a deterministic cascade model on network $\G$. Each node is endowed with a threshold value $\alpha_v\in[0,1]$. 
At time $t=0$ a subset of nodes $S_0$ is chosen to be the seed set, representing the initial EV-adopters. Letting $S_t$ be the set of new EV-adopters at time $t$ and the set $ S_t^{\star}:=\cup_{\tau=0}^{t}S_{\tau}$ represent the total amount EV-adopters in the network until time $t$, for each discrete time $t\geq1$ one has 
\begin{equation}\label{eq:dynamics}
S_{t}=\left\{v\in\V\setminus (\cup_{\tau=0}^{t-1}S_{\tau}):\frac{|S_t^{\star}\cap\N_v|}{|\N_v|}\right\}\geq \alpha_v.
\end{equation}
The dynamics in \eqref{eq:dynamics} describes a model of EV diffusion where the adoption of the agents depends on the relative popularity of EVs among neighbors. 

It should be noticed that $S_t^{\star}$ is monotonically increasing in $t\in\mathbb{N}$ and the dynamics converge to a final adopters set. In \cite{6160999} the final adopter set is characterized in terms of the network, seed set and threshold values making use of the concept of cohesive sets, the definition of which is as follows.

\begin{definition}[Cohesive set] A set $\Omega\subseteq \V$ is said cohesive if for all $\omega\in\Omega$
$$
\frac{|\Omega\cap\N_{\omega}|}{|\N_{\omega}|}>1-\alpha_{\omega}.
$$
\end{definition}

This definition states that a set $\Omega$ is cohesive if for each element $\omega$ in the set the ratio of neighbors which do not belong $\Omega$ is strictly smaller than the value of the threshold $\alpha_{\omega}$.

\begin{theorem}[Lemma 2 in \cite{6160999}]
Given a network with seed set $S_0\subset\V$, let $\Omega\subset\V\setminus S_0$ be the cohesive set with maximal cardinality. Then the dynamics converges in finite time, and the set of final adopters is  given by $\overline{S}^{\star}=\V\setminus\Omega
$.
\end{theorem}

Computing the maximal cohesive set contained in the complement of the seed set is computationally expensive and an algorithm, based on linear programming methods, is proposed in \cite{ROSA2013322}.

It is worth remarking that our goal is to study the spread of adoption in a finite time horizon.
More precisely, in next section the cascade EV-adoption model is simulated for $t\in [0,T]$, using a network topology and features of the model that are directly inferred by real mobility patterns, as will be better explained in the next section. 

\section{Data-based description of the EV adoption process}\label{sec:adoption}
To build the network described before, we use a set comprising data of $1,000$ vehicles registered in the Italian province of Parma. The available dataset includes anonymized GPS traces collected over one year (from $1$st September 2017 to $31$st August 2018) from conventional ICE vehicles equipped with telematic e-Boxes. Each record contains information on the instantaneous GPS latitude and longitude related to an \emph{event} (such as vehicle ignition or shutdown), along with their associated time stamp. By properly aggregating the available information, \emph{trips} and \emph{stops} can be extracted from these records, along with their duration and the traveled distance, thus allowing us to infer the mobility pattern of each vehicle. The reader is referred to \cite{zinnari} for further details on the dataset\footnote{The $1,000$ vehicles used in this study are randomly selected among the $21,251$ available ones, so to maintain the statistics on EV suitability presented in \cite{zinnari}.}.

When studying the EV-adoption process, it seems reasonable to assume that one's opinion can be influenced by the point of views of its \emph{closest} neighbors. This models that fact that actually seeing neighbors buying new EVs can help fostering individuals' inclination to do a similar choice. In this light, EV adoption is investigated based on the proximity between the \emph{homes} of the agents, represented by those stops that take place mostly overnight and that are sufficiently long. For each agent, the GPS coordinates of its home are retrieved from the corresponding records as the centroids of the clusters of proximal stops that contain at least $50$ \% of the overnight stops. In turn, the latter are empirically detected as \emph{long} stops (duration above $7$ hours) that take place (at least partially) during night hours. 

More formally, $h_{v}=(long_{v},lat_{v})$ be the GPS coordinate associated to the home of the $v$-th agent, with $v \in \V$, and denote as $d(h_{v},h_{w})$~[km] the geodesic distance between the homes of the $v$-th and $w$-th agents. For the proximity-based study to be well grounded, the initial dataset is prepossessed to exclude all vehicles $v \in \V$ such that
\begin{equation}
d(h_{v},h_{w})>60~\mbox{[km]},~~ \forall~ w \in \V, ~w \neq v,
\end{equation}
namely we neglect the agents whose home is not \emph{sufficiently} close to the one of all the others. This leads to a minor reduction in the number of agents considered in our study, with the preprocessed dataset comprising the records of $994$ vehicles. 
	\begin{figure}[!tb]
		\centering
		\includegraphics[scale=.7]{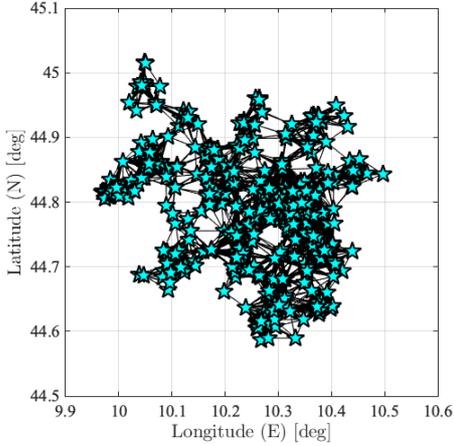}
		\caption{Network topology based on proximity measures. Cyan stars indicate the position of the homes, while black lines denote edges between neighbor nodes.}\label{Fig:Graphwithoutclasses}
	\end{figure}
\subsection{Network extraction}\label{sec:Network_extraction}
Let us assume that the homes of neighboring agents are proximal if they are separated by less than $d_{max}=5$ [km]. We stress here that, in extracting the data-driven network to investigate EV adoption, we aim at obtaining a \emph{connected} graph. 
%This guarantees that every agent can eventually change its opinion with respect to EV adoption.

\begin{figure}[!tb]
		\centering
		\includegraphics[width=0.95\columnwidth]{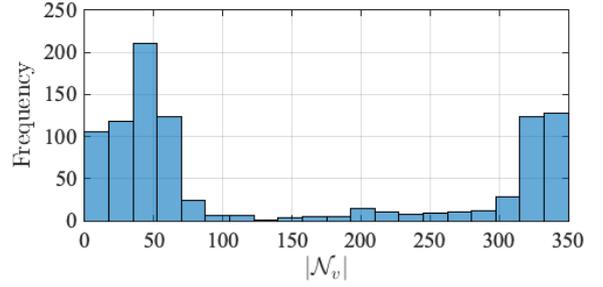}
		\caption{Degree distribution for the proximity-based network $\G$.}\label{Fig:DegreeHist}
	\end{figure}

By following the rationale dictated by home proximity, we construct a preliminary network $\G=(\V,\E)$, such that
\begin{equation}\label{eq:first_cond}
(v,w) \in \E \iff d(h_{v},h_{w}) \leq 5~\mbox{[km]}.
\end{equation}
However, the graph $\G$ does not result to be \emph{connected}, thus entailing the presence of groups of agents with intra-cluster distance verifying the condition in \eqref{eq:first_cond}, but which are actually detached from the other nodes. Since we want to obtain a connected network, we prune $\G$ by progressively removing the agents that fulfill the following conditions. 
\begin{itemize}
	\item \emph{Remote agents}, whose home is separated from all the others by more than $d_{max}$~km, namely $v \in \V$ such that
	\begin{equation}
	d(h_{v},h_{w})>5~\mbox{[km]},~~ \forall w \in \V,~w \neq v. 
	\end{equation} 
	\item  \emph{Outliers}, namely agents with homes that are distant from the average house position. Let $\mu_{h}=(long_{\mu},lat_{mu})$ be the point whose GPS coordinates are given by
	\begin{equation}
	long_{\mu}=\frac{1}{|\V|} \sum_{v \in \V} long_{v},~~~ lat_{\mu}=\frac{1}{|\V|} \sum_{v \in \V} lat_{v},
	\end{equation}
	and denote with $\sigma_{h}$ the standard deviation of the distances between the agents' homes and $\mu_{h}$. In this work, we classify a vehicle $v \in \V$ as outlier if
	\begin{equation}
	d(h_{v},\mu_{h})>\sigma_{h}~\mbox{[km]}.
	\end{equation}  
	\item \emph{Almost isolated agents}, \emph{i.e.,} the nodes such that
	\begin{equation}
	|\N_{v}|<8, ~~ v \in \V.
	\end{equation}
\end{itemize} 
Note that, by trimming almost isolated nodes, we implicitly assume that agents with few neighbors are unlikely to change their opinion on EV adoption. As a result of this pruning procedure, we obtain the \emph{connected} graph shown in \figurename{\ref{Fig:Graphwithoutclasses}}, which comprises $894$ vehicles (\emph{i.e.,} approximately 89\% of the agents available after preprocessing). The associated distribution of the connection degree is shown in \figurename{\ref{Fig:DegreeHist}}, showing that most of the nodes have a relatively low degree, but there is also a fair number of agents that are regarded as proximal to $38$\% of the overall nodes.  
\subsection{Data-driven agents classification}
	\begin{figure*}[!tb]
		\centering
		\begin{tabular}{cc}
		\subfigure[EV adoption categorization.\label{Fig:Categorization}]{\scalebox{.8}{\begin{tikzpicture}[]
		\foreach \percent/\name in {
			29.5/Initial EV-adopters,
			70.5/Others
		} {
			\ifx\percent\empty\else               % If \percent is empty, do nothing
			\global\advance\cyclecount by 1     % Advance cyclecount
			\global\advance\ind by 1            % Advance list index
			\ifnum3<\cyclecount                 % If cyclecount is larger than list
			\global\cyclecount=0              %   reset cyclecount and
			\global\ind=0                     %   reset list index
			\fi
			\pgfmathparse{\cyclelist[\the\ind]} % Get color from cycle list
			\edef\color{\pgfmathresult}         %   and store as \color
			% Draw angle and set labels
			\draw[fill={\color!40},draw={\color}] (0,0) -- (\angle:\radius)
			arc (\angle:\angle+\percent*3.6:\radius) -- cycle;
			\node at (\angle+0.5*\percent*3.6:0.7*\radius) {\percent\,\%};
			\node[pin=\angle+0.5*\percent*3.6:\name]
			at (\angle+0.5*\percent*3.6:\radius) {};
			\pgfmathparse{\angle+\percent*3.6}  % Advance angle
			\xdef\angle{\pgfmathresult}         %   and store in \angle
			\fi
		};
		\end{tikzpicture}}} &
		\subfigure[Graph with labeled nodes.\label{Fig:GraphwithClasses}]{\includegraphics[scale=.7]{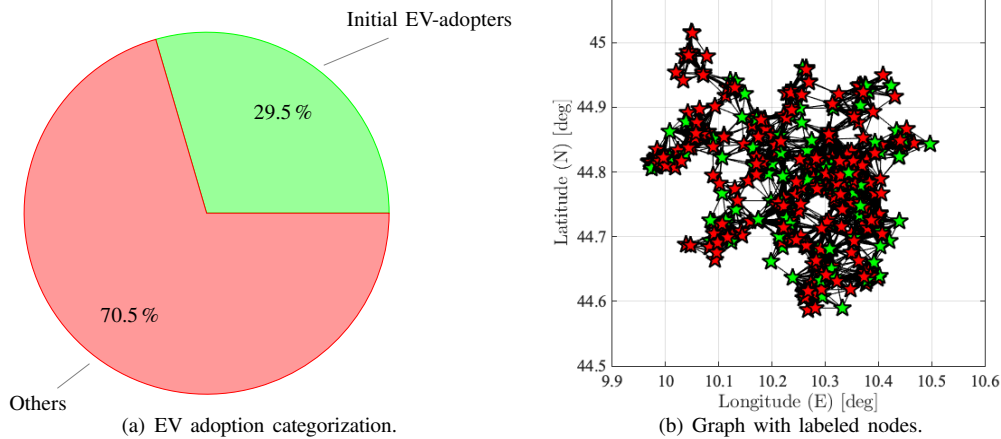}}
		\end{tabular}
		\caption{Agent classification and labeled network. Green nodes represent the initial EV-adopters $S_0$.}
	\end{figure*}
To select the seeds $S_{0}$, representing those agents who are assumed to actually having switched to an EV already, we rely on information on the electrification potential of each agent to EV adoption based on the inferred individual motion patterns, by exploiting the rationale in \cite{zinnari}. Accordingly, the suitability to EV adoption is defined based on the characteristics of the daily trips performed by each agent, whenever the corresponding vehicle is used. In particular, the agents are classified according to the distance covered each day, under reasonable assupmtions on both the driver's behavior and the technological limitations of EVs, \emph{e.g.,} the EV range (without recharge) is fixed at 200 km, see again \cite{zinnari}. 

By restricting ourselves to two classes only (adopters and non-adopters), initially one and only one of the following labels can be associated to each agent:
\begin{itemize}
	\item\emph{perfectly suitable for EV adoption}, if the daily trips of the agent never exceed $200$ km;
	\item \emph{not immediately suitable EV adoption}, if the daily trips of the agent exceed $200$ km, independently of the possibility of recharge linked to the individual motion patterns.
\end{itemize} 
In this paper, these labels are exploited as a proxy for the personal inclination towards EVs and their adoption. Indeed, it seems likely that people able to perform their daily trips without facing the problems related to the limitations of EVs would be more inclined to adopt this new technology. As a consequence, in principle $S_{0}$ is constituted by all agents that, at time $t=0$, are associated to the first label. By categorizing the nodes, the graph in \figurename{\ref{Fig:Graphwithoutclasses}} becomes the one reported in \figurename{\ref{Fig:GraphwithClasses}}, which shows how initial EV-adopting agents are positioned within the network. Additional insights on the dimension of the seed set is provided in \figurename{\ref{Fig:Categorization}}, which shows that $29.5$\% of vehicles belong to $S_{0}$, \emph{i.e.,} $264$ out of $894$ nodes of the data-driven network.

\section{Data-driven network analysis}\label{sec:anal}
\begin{figure*}[!tb]
	\centering
	\begin{tabular}{ccc}
		\subfigure[Initial configuration \label{Fig:initial_configuration}]{\includegraphics[scale=.6]{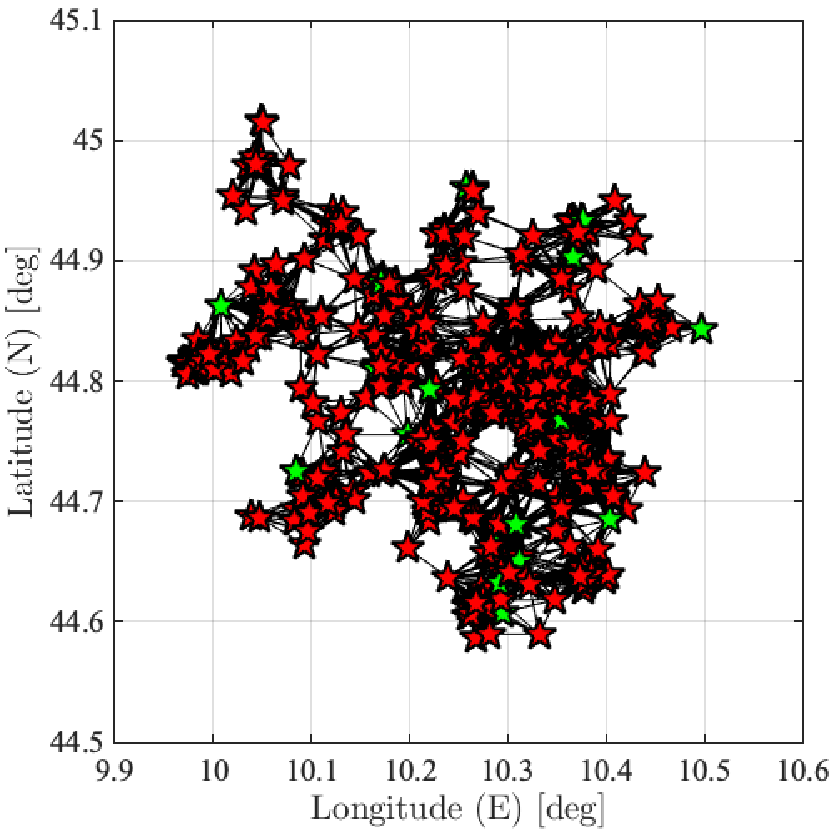}} & \subfigure[After around 2 years]{\includegraphics[scale=.6]{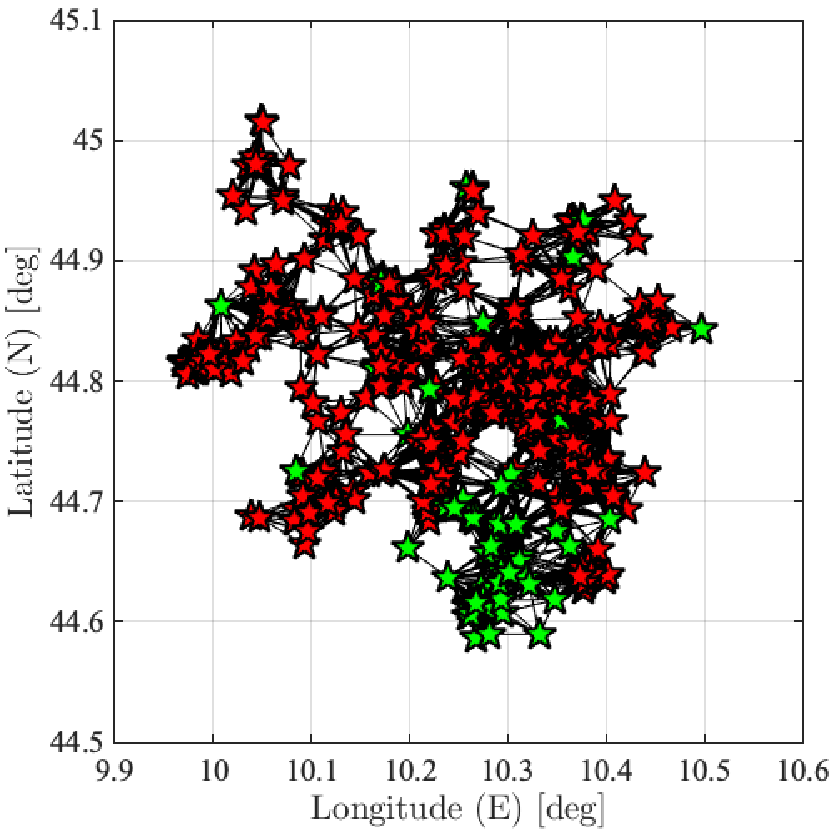}} & \subfigure[After 5 years]{\includegraphics[scale=.6]{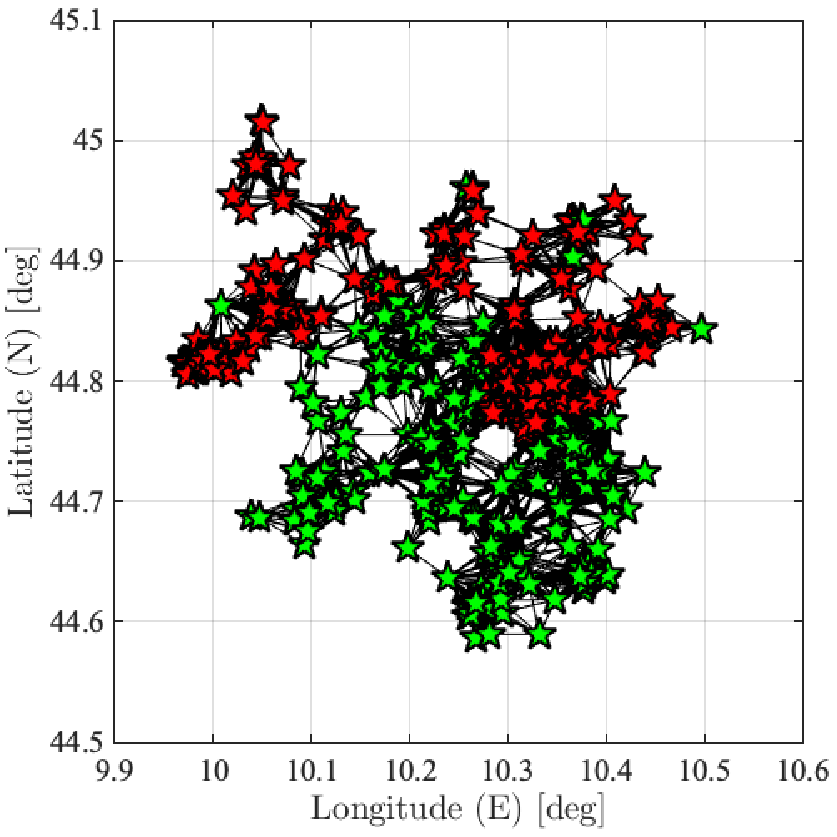}}
	\end{tabular}
	\caption{Typical case: opinion diffusion over time on the network.}\label{Fig:typicalEvolution}
\end{figure*}

In this section we consider a few scenarios of interest to evaluate the sensitivity of the EV diffusion model described in \eqref{eq:dynamics} to different parameters. For our experiments to be consistent with the actual percentage of EVs registration in Italy (namely, $7$ \% of the overall vehicle registrations), $S_0$ is chosen uniformly at random among the drivers that have been classified as perfectly suitable for EV adoption, instead of making the seed set coincide with the cluster $\mathcal{C}_{1}$ of vehicles classified as \emph{perfectly suitable for EV adoption}. Specifically, we fix $|S_0|=\eta |\mathcal{C}_1|$, with $\eta=0.25$, and we perform 20 tests by randomly re-sampling $\mathcal{C}_{1}$ so as to consider different initial configurations. A specific instance of the network at $t=0$ is shown in \figurename{~\ref{Fig:initial_configuration}}, that clearly shows the limited number of seeds with respect to the overall number of agents.
\begin{figure}[!tb]
	\centering
	\includegraphics[width=0.99\columnwidth]{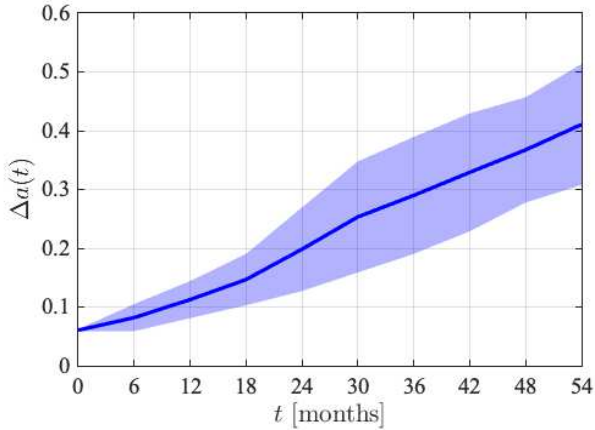}
	\caption{Mean and standard deviation of the percentage EV-adopters vs time.\label{Fig:mean_time}}
\end{figure}
Initially, we set the parameters $\alpha_v$ to be all equal to $\alpha=0.15$ for all $v\in\V\setminus S_0$. \figurename{\ref{Fig:mean_time}} shows the average percentage $\Delta a$ [\%] of EV-adopters (averaged over 20 instances) as a function of time $t~\in~[0,54]$. In our experiments, each time step corresponds to six months, so that we evaluate EV adoption over a span of around five years, which seems a reasonable assumption considering that technological changes on longer time spans might change the agents perception on EV adoption. Despite the relatively low number of seeds, the overall percentage of EV-adopters increases over time up to $40\%$. 

\begin{figure}[!tb]
	\centering
	\includegraphics[width=0.99\columnwidth]{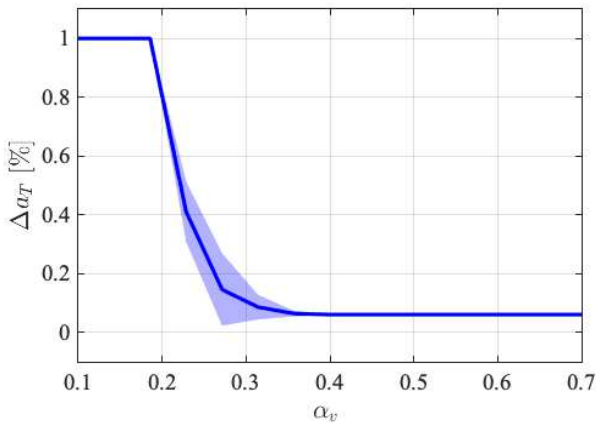}
	\caption{Mean and standard deviation of the percentage EV-adopters after 5 years vs threshold $\alpha_{v}$.\label{Fig:mean_alpha}}
\end{figure}

In a second experiment, we analyze the sensitivity of the model to changes in the threshold value. Specifically, we consider increasing values of $\alpha_{v}$ within the interval $\alpha_{v}\in[0.01,0.7]$, so as to evaluate both settings in which the agents are keen on adopting EVs, \emph{i.e.,} when $\alpha_{v}=0.1$ quite few neighbors ($10$\%) must have adopted EVs for an agent to adopt an electric vehicle itself, to situations in which the agents are generally very skeptical about EVs, namely when $\alpha_{v}=0.7$. The average percentage of EV diffusion (over 20 instances) after 5 years is shown in \figurename{\ref{Fig:mean_alpha}} as a function of the threshold. It is clear that the more the agent are not convinced by electric vehicles, the least the final percentage of adopters is. It is interesting to note that the values of $\alpha_{v}$ in the range $\alpha_{v}\in[0.15,0.4]$ seem to be critical for EV adoption. Indeed, for $\alpha_{v} \in (0.15,\ 0.4)$ the percentage of final adopters sensibly decreases and, for $\alpha_{v} \geq 0.4$ the opinion does not generally diffuses throughout the network.

\begin{figure}[!tb]
	\centering
	\includegraphics[width=0.9\columnwidth]{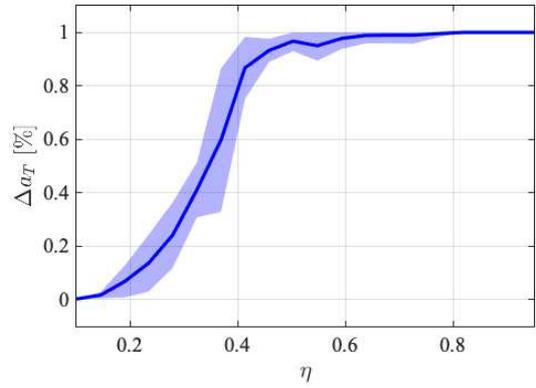}
	\caption{Mean and standard deviation of the percentage of EV-adopters after 5 years vs percentage of seeds $\eta$ with respect to the number of agents perfectly suitable for EVs.\label{Fig:mean_eta}}
\end{figure}

As a final test, we consider how EV diffusion is affected by different choices of $\eta$, namely the number of initial EV-adopters, when $\alpha_v=0.15$. Note that, for $\eta=0.25$, the choice of this threshold corresponds to a setting in which we get close to reaching consensus over the network within the considered steps (see \figurename{~\ref{Fig:mean_alpha}}). The final percentage of EV-adopters obtained for $\eta=[0.1,\ 0.95]$ is reported in Figure \ref{Fig:mean_eta}. It is clear that it is sufficient to have approximately $15$\% of the perfectly suitable agents as seeds to trigger a cascade adoption over the network. 

By looking at Figures~\ref{Fig:mean_time}-\ref{Fig:mean_eta}, the expected percentage of diffusion is concentrated around the average value for different random simulation. By considering the initial instance reported in \figurename{~\ref{Fig:initial_configuration}}, which is obtained by fixing $\eta=0.25$, in \figurename{~\ref{Fig:typicalEvolution}} we show an instance of the obtained opinion diffusion process. Note that, in this specific case study, we still consider a relatively low number of initial EV-adopters, as this reflects the quite realistic scenario in which even drivers that are perfectly suited for EVs as far as driving habits are concerned, do not adopt EVs also due to the absence of social influences. As it can be seen, although few drivers have adopted EVs, after almost 5 years of evolution, the EV adoption diffuses to quite a large portion of the network, while a relatively small set of agents remains resistant to the change. Note that the agents not adopting EVs are those belonging to regions with a very low number of initial seeds, which seems to be too isolated to actually influence their surroundings.

\begin{figure}[!tb]
	\centering
	\begin{tabular}{cc}
	\subfigure[Random sampling \label{Fig:incentives_random}]{\includegraphics[scale=.9]{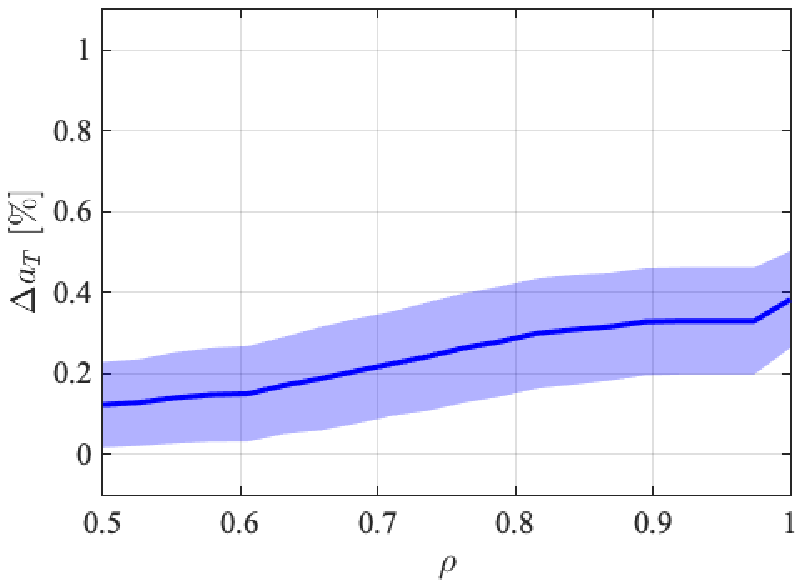}} \\ 	\subfigure[Degree-based sampling\label{Fig:incentives_grade}]{\includegraphics[scale=.9]{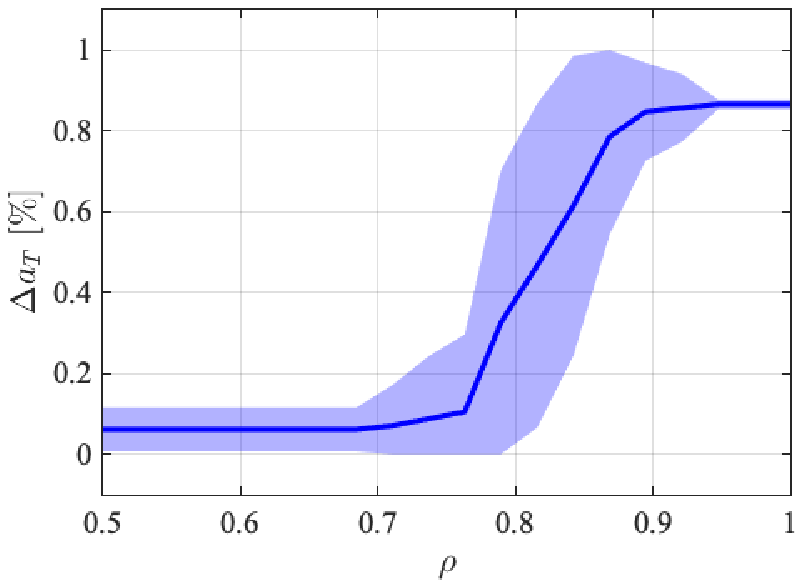}}
	\end{tabular}
\caption{Mean and standard deviation of the adoption rate $\Delta a_{T}$ after $5$ years vs $\rho$, with $\alpha=0.15$ and $\eta=0.10$.}\label{Fig:Incentives_strategies}
\end{figure}

\section{Incentive Policy}\label{sec:incentive}
Based on the analysis on the sensitivity to parametric changes in the model, we now consider how to assess the effect of providing incentives on the spread of EV-adoption. Our main goal here is to identify, if possible, a set of influential nodes in the network that can guide the choices of the other agents, thus boosting EVs acceptance. Note that, although in this preliminary analysis we study the effect of incentives empirically, this problem is strictly related to the influence maximization problem over cascading networks \cite{Lim16}, that can be summarized as follows. 
\begin{problem}[Influence maximization problem \cite{ROSA2013322}] 
 Find the set $\Gamma\in \V\setminus S_0$ such that $|\Gamma| = k$ and the number $S^{\star}_T$ of adopters until a finite horizon $T$ is maximal.  
\end{problem}
This problem is challenging and computationally expensive given its combinatorial nature, since the number of possible configurations of the thresholds is of order ${O}(n^k/k^k)$.

Consider the proximity network $\G = (\V,\E)$ extracted in Section \ref{sec:Network_extraction}. As in the previous section, assume that at time $t=0$ the set of initial EV-adopters $S_0$  is chosen uniformly at random among the drivers that are perfectly suitable for EV adoption, so that $|S_0|/|\mathcal{C}_{1}|=\eta$ with $\eta=0.25$. Let $\Gamma\subseteq\V\setminus S_0$ and consider the dynamics described in \eqref{eq:dynamics} with 
\begin{equation}
\alpha_v=\begin{cases}
(1-\rho)
\alpha& \forall v\in\Gamma\\
\alpha& \forall v\notin\Gamma\cup S_0
\end{cases}
\end{equation}
The parameter $\rho \in [0.5, \ 1]$ provides indications on the incentive that has been given by a third party, \emph{e.g.,} governments, policy or car makers, to favor EV adoption. Indeed, if the we increase the magnitude of parameter $\rho$, we reduce the threshold associated to the agent $v\in\Gamma$, which will in turn be more willing to adopt the EV. Initially, we consider a greedy approach to construct the set of nodes $\Gamma$ to which incentives are given. In particular, we select a fraction of potential users uniformly at random so that $|\Gamma|/|\V-S_0|=0.15$ (for  $\eta=0.1$ and $\alpha=0.15$). The final adoption percentage after 5 years is reported in \figurename{~\ref{Fig:incentives_random}} as a function of $\rho$, with the thick curve representing the average percentage of adoption after 5 years (over 100 instances). This result shows that the considered incentive policy generally leads to a relatively small improvement in the adoption of electric vehicles. To exploit the information that can be drawn from the proximity network, we then find the set $\Gamma$ among the nodes with the highest degree, which is regarded here as a proxy for the centrality of each node. By looking at \figurename{\ref{Fig:incentives_grade}}, it is clear that even a simple incentive strategy based on the degree of centrality \cite{Newman:2010:NI:1809753} is able to guarantee a large diffusion. The result of this study shows that the targeting strategy plays a key role to boost the adoption process of EVs. Indeed, while the percentage of adoption is limited to the range $[0, \ 0.4]$ if incentives are given to randomly chosen agents, the use of degree in choosing the agents to be supported allows to have a large gain, leading to final $0.8$ percentage of diffusion, with fixed network topology.   

\section{Concluding remarks}
In this paper, a network-based modelling of mobility patterns has been presented, that allows us to model the proximity among different users within a graph-based framework. Based on this, the social network of interest was built based on the description of real mobility data measured from telematic e-boxes, considering how close the estimate home locations of the different users are. This network was then used to setup an analysis of the adoption an electric vehicles from users, starting from the initial suitability to electric mobility that was extracted from the initial data. The opinion diffusion patterns have been analysed under different scenarios, and an incentive policy has been outlined to foster the adoption rate.
Ongoing work is being devoted to further investigate the potential of the adopted framework, considering other proximity measures and using also social variables, such as, for example, the users' income, to refine the adoption mechanism.

\bibliographystyle{IEEEtran}

\bibliography{paper_EV_social_THMS_conf_final.bib}

\begin{thebibliography}{10}
\providecommand{\url}[1]{#1}
\csname url@rmstyle\endcsname
\providecommand{\newblock}{\relax}
\providecommand{\bibinfo}[2]{#2}
\providecommand\BIBentrySTDinterwordspacing{\spaceskip=0pt\relax}
\providecommand\BIBentryALTinterwordstretchfactor{4}
\providecommand\BIBentryALTinterwordspacing{\spaceskip=\fontdimen2\font plus
\BIBentryALTinterwordstretchfactor\fontdimen3\font minus
  \fontdimen4\font\relax}
\providecommand\BIBforeignlanguage[2]{{%
\expandafter\ifx\csname l@#1\endcsname\relax
\typeout{** WARNING: IEEEtran.bst: No hyphenation pattern has been}%
\typeout{** loaded for the language `#1'. Using the pattern for}%
\typeout{** the default language instead.}%
\else
\language=\csname l@#1\endcsname
\fi
#2}}

\bibitem{docherty2018governance}
I.~Docherty, G.~Marsden, and J.~Anable, ``The governance of smart mobility,''
  \emph{Transportation Research Part A: Policy and Practice}, vol. 115, pp.
  114--125, 2018.

\bibitem{grelier_2018}
F.~Grelier, \emph{CO2 EMISSIONS FROM CARS: the facts}, 2018.

\bibitem{efficiency2017}
{U.S. Department of Energy - Office of Energy Efficiency and Renewable energy},
  \emph{Electric-Drive Vehicles}, 2017,
  https://afdc.energy.gov/files/u/publication/electric\_vehicles.pdf.

\bibitem{needell2016potential}
Z.~A. Needell, J.~McNerney, M.~T. Chang, and J.~E. Trancik, ``Potential for
  widespread electrification of personal vehicle travel in the united states,''
  \emph{Nature Energy}, vol.~1, no.~9, p. 16112, 2016.

\bibitem{barr2014smarter}
S.~Barr and J.~Prillwitz, ``A smarter choice? exploring the behaviour change
  agenda for environmentally sustainable mobility,'' \emph{Environment and
  Planning C: government and policy}, vol.~32, no.~1, pp. 1--19, 2014.

\bibitem{globalEV2019}
{International Energy Agency}, \emph{Global EV Outlook 2019 - Scaling-up the
  transition to electric mobility}, 2019.

\bibitem{mckinsey_2017}
{McKinsey \& Company}, \emph{Electrifying insights: How automakers can drive
  electrified vehicle sales and profitability}, 2017.

\bibitem{6160999}
D.~{Acemoglu}, A.~{Ozdaglar}, and E.~{Yildiz}, ``Diffusion of innovations in
  social networks,'' in \emph{2011 50th IEEE Conference on Decision and Control
  and European Control Conference}, 2011.

\bibitem{frasca2013gossips}
P.~Frasca, C.~Ravazzi, R.~Tempo, and H.~Ishii, ``Gossips and prejudices:
  Ergodic randomized dynamics in social networks,'' \emph{IFAC Proceedings
  Volumes}, vol.~46, no.~27, pp. 212--219, 2013.

\bibitem{friedkin2016network}
N.~E. Friedkin, A.~V. Proskurnikov, R.~Tempo, and S.~E. Parsegov, ``Network
  science on belief system dynamics under logic constraints,'' \emph{Science},
  vol. 354, no. 6310, pp. 321--326, 2016.

\bibitem{lorenz2007continuous}
J.~Lorenz, ``Continuous opinion dynamics under bounded confidence: A survey,''
  \emph{International Journal of Modern Physics C}, vol.~18, no.~12, pp.
  1819--1838, 2007.

\bibitem{ROSA2013322}
D.~Rosa and A.~Giua, ``On the spread of innovation in social networks,''
  \emph{IFAC Proceedings Volumes}, vol.~46, no.~27, pp. 322 -- 327, 2013, 4th
  IFAC Workshop on Distributed Estimation and Control in Networked Systems
  (2013).

\bibitem{zinnari}
Z.~F., S.~S., T.~M., F.~S., S.~G., and S.~S.M., ``Mining the electrification
  potential of fuel-based vehicles mobility patterns: a data-based approach,''
  in \emph{First IEEE Conference on Human-Machine Systems}, 2020, submitted.

\bibitem{Lim16}
Y.~Lim, A.~Ozdaglar, and A.~Teytelboym, ``A simple model of cascades in
  networks,'' \emph{submitted}, 2016.

\bibitem{Newman:2010:NI:1809753}
M.~Newman, \emph{Networks: An Introduction}.\hskip 1em plus 0.5em minus
  0.4em\relax New York, NY, USA: Oxford University Press, Inc., 2010.

\end{thebibliography}

\end{document}